\documentclass{elsart}
\usepackage{graphicx}
\begin{document}

\begin{frontmatter}
\title{From fracture to fragmentation: discrete element modeling - Complexity of crackling noise and fragmentation phenomena revealed by 
discrete element simulations} 

\author{Humberto A. Carmona}
\author[2]{Falk K. Wittel} 
\author[3]{Ferenc Kun} 

\address{Departamento de F\'\i sica, Universidade Federal do Cear\'a, 60451-970 Fortaleza, \\ Cear\'a, Brazil }
\address[2]{Computational Physics, IfB, ETH Z\"urich, Stefano-Franscini-Platz 3, CH-8093 Z\"urich, } 
\address[3]{Department of Theoretical Physics, University of Debrecen, H-4010 Debrecen, P.O.Box: 5, Hungary} 

 \begin{abstract}
Discrete element modelling (DEM) is one of the most efficient computational 
approaches to the fracture processes of heterogeneous materials on mesoscopic 
scales. From the dynamics of single crack propagation through the statistics of 
crack ensembles to the rapid fragmentation of materials DEM had a substantial 
contribution to our understanding over the past decades. Recently, the 
combination of DEM with other simulation techniques like Finite Element 
Modelling further extended the field of applicability. In this paper we briefly 
review the motivations and basic idea behind the DEM approach to cohesive 
particulate matter and then we give an overview of on-going developments and 
applications of the method focusing on two fields where recent success has been 
achieved. We discuss current challenges of this rapidly evolving field and 
outline possible future perspectives and debates.
\end{abstract}
\end{frontmatter}
\newpage

\section{Introduction} \label{intro}
Fracture and fragmentation of materials have been the subject to 
human interest for as long as we can think, mainly due to practical reasons. 
For 
centuries research on fracture was mainly driven by catastrophic 
failure events of engineering constructions which occurred due to the poor 
understanding 
of relevant processes \cite{herrmann,eliz,mikko}. 
Names like da Vinci, Galilei, Griffith, Weibull, W\"ohler, and Inglis
among others 
are all related to engineering solutions to fracture \cite{eliz}.
The nature of fracture phenomena, however, impeded 
systematic theoretical studies. No more than three decades ago mainstream 
physics slowly started to study fracture and fragmentation problems, driven by 
the discoveries of a young generation of researchers that made computers 
accessible for their research \cite{herrmann,mikko}. 
Based on the pioneering works of Hans J.\ Herrmann and others, lattice models, fuse 
models and meshless particle models emerged for fracture studies that - driven by the 
breath-taking advances of computational and algorithmic capabilities - proved 
to be very successful for studying fracture and fragmentation phenomena 
\cite{herrmann,eliz,mikko}. Around that time, Cundall proposed a particle 
method with rigid body dynamics to model fracture of frictional cohesive materials, 
characteristic of geotechnical applications \cite{dem_basic_1}. Under the name 
Discrete Element Method (DEM) a group of approaches emerged that generate the 
motion of an assembly of particles starting from the dynamics of its 
constituents. The similarities of the DEM to popular methods in other fields of 
research like molecular dynamics \cite{allen_til} or smooth particle dynamics 
\cite{sph}, lead to cross-fertilization in algorithmic development. Today DEM 
is 
a powerful tool to simulate the breaking of heterogeneous materials beyond the 
point of single crack growth. Various particle geometries, material response, 
ways to treat cohesive, repulsive behavior and of course loss of cohesion lead 
to a flexible tool-set of approaches. Strategies for higher order 
agglomeration, 
coupling to continuum domains or particle based fluid solvers like lattice 
Boltzmann extended the reach of DEM significantly \cite{F1,F2,F3,F4,F5}. Today, 
applications of DEM made a substantial contribution to the understanding of the 
mechanical response and breaking phenomena of heterogeneous materials under 
various types of loading conditions. Ranging from the slowly changing 
sub-critical loads to the highly energetic fragmentation, DEM proved to be an 
indispensable tool for investigations.  

In this article we briefly review the motivations and basic ideas behind the 
DEM 
approach, as well as, its current extension by coupling to a continuum domain. 
Among the widespread applications of DEM for the fracture of heterogeneous 
materials we highlight two fields where recently DEM have played a decisive 
role 
to achieve major success. Finally we discuss remaining challenges of this 
rapidly evolving field and outline possible future perspectives.

\section{Discrete Element Models for cohesive particulate materials} 
\label{sec:1}
Fracture and fragmentation are difficult problems to handle numerically due to 
the continuous generation and evolution of crack surfaces. Classical numerical 
methods such as Finite Element, Finite Differences, or Boundary Elements solve 
partial differential equations of continuum mechanics, so that they are able to 
consider only a small number of discontinuities and cannot encompass the entire 
fracturing process. Discrete Element Modelling (DEM) is a computational 
approach 
to the deformation and failure of cohesive frictional materials which embeds 
materials' complexity  by representing it with a set of discrete elements. The 
method is physically based in the sense that the elements of discretization are 
physical entities having e.g.\ mass and velocity, hence, they are called 
particles. 
The interaction of particles is defined such that the model accounts for the 
proper macroscopic response of the medium including both, constitutive behavior 
and  failure mechanisms. The approach was introduced by Cundall and Strack 
\cite{dem_basic_1} in 1979 which then initiated a rapid development of the 
technique and a wide variety of applications in diverse fields of engineering, 
physics, and geosciences 
\cite{dem_basic_1,dem_basic_2,dem_basic_3,dem_basic_4,dem_basic_5,kun_basquin}. 

\subsection{Model construction}\label{sec:model}
DEM is best suited for materials which are inherently disordered on the 
mesoscopic scale, i.e.\ they are composed of grains of various shapes with 
complicated cohesive coupling in between. To begin with, the model has to give 
a 
high quality representation of the microstructure of the specific material 
considered. To keep the problem numerically tractable, particle shapes are 
usually idealized by spheres in three dimensions (3D) so that the problem of 
discretization is reduced to the generation of a random homogeneous packing of 
spherical particles with a prescribed size distribution and a desired density. 
Two classes of generation methods can be distinguished, i.e.\ the dynamic and 
constructive methods \cite{geomalg,bagi1,donze1,donze2,feng1,cui1}: Dynamic 
methods typically start from a random configuration of point-like particles 
which are then gradually blown up to the desired size reaching a dense 
arrangement in the domain of interest. As an alternative, particles with the 
required extension can be placed in a volume significantly larger than the 
domain of interest and then either the volume can be slowly reduced until an 
appropriate packing is reached or the particle system can be compactified under 
the action of a force field \cite{bagi1,dem_basic_3,humb1,kun_rup_1,kun_rup_2}. 
 
Already the packing generation involves demanding simulations of the motion of 
particles making dynamical methods rather time consuming. Constructive 
algorithms take a different strategy, namely, they are purely based on 
geometrical procedures to discretize the spatial domain in terms of spheres 
\cite{donze1,cui1}. Efficient algorithms have been developed to fill containers 
of various shapes which all share the feature that they lead to packings with a 
low coordination number and a high porosity. The density can be increased by 
using the sedimentation technique \cite{bagi1,geomalg,donze1} or by gradual 
refinement of the packing using tetrahedral meshes \cite{donze1}. 
\begin{figure}
\centering{\includegraphics[scale=0.7]{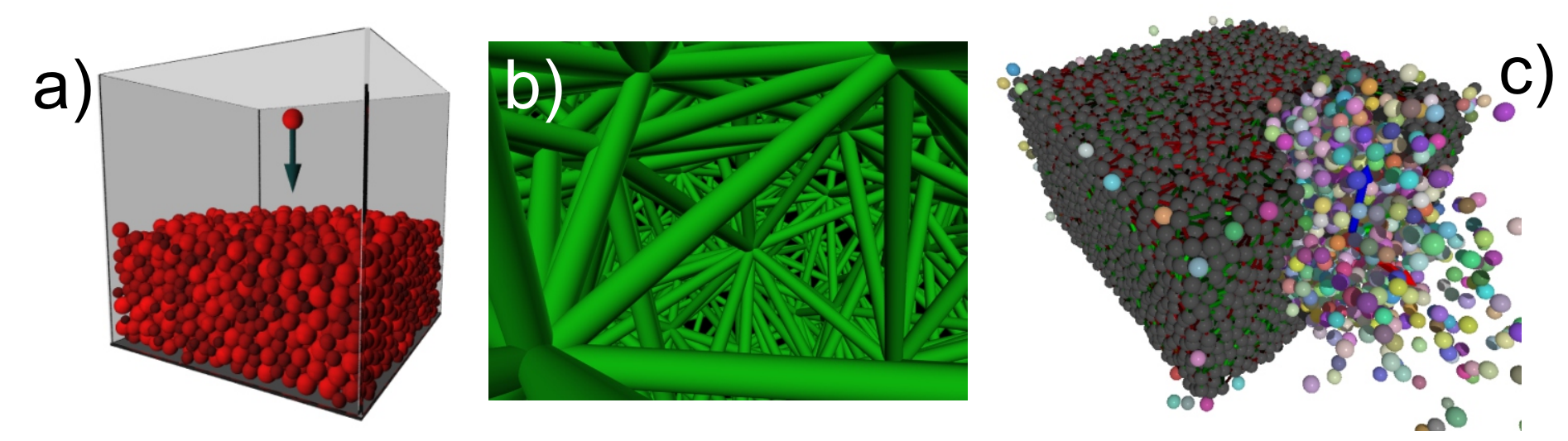}}
 \caption{Demonstration of the construction of DEMs: $(a)$ Random homogeneous 
packing of spherical particles is generated by particle deposition in a 
rectangular container \cite{kun_rup_1}.  $(b)$ Delaunay tetrahedral mesh is 
constructed with the particle centers and beams are introduced along the edges 
of tetrahedra. A close-up on the beam lattice is presented in 
\cite{kun_rup_1,humb1}.  $(c)$ An early stage of the impact fragmentation of a 
rectangular sample generated by a projectile which hits the middle of the front 
side of the body. Beams are colored according to the axial strain where 
stretched and compressed beams have red and green colors, respectively. Colors 
are randomly assigned to fragments. \label{fig:dem}}
\end{figure} 

Under external load, the particle ensemble deforms and cracks emerge at highly 
stressed locations, the physics of which has to be captured by the dynamics of 
inter-particle contacts. Since the numerical representation of the deformation 
of particles is computationally not feasible, contact models rely on the 
overlap 
of the spherical particles and express the normal component of the contact 
force 
in terms of the overlap distance. In this so-called soft particle contact 
model, 
tangential forces and torques depend on the relative displacement of the 
particles since contact has been established. Realistic contact models capture 
dissipation, rolling and torsion resistance, as well as elastic-plastic contact 
deformation \cite{luding1,thornton,humb1,donze2,grandynbook}. 

Cohesion of the material arising due to bonding of its grains can be captured 
by 
coupling neighboring particles via elastic spring or beam elements. In the most 
realistic case beams in 3D account for the stretching, compression, shear, 
bending, and torsion of cohesive contacts \cite{dem_basic_3,grandynbook,humb1}. 
Beam elements may act as bonds coupling either the surfaces \cite{dem_basic_3} 
or the centers of mass of particles 
\cite{dem_basic_4,dem_basic_5,humb1,kun_rup_1}. The geometrical features of 
beams, i.e.\ length, cross-sectional area, and moments are determined by the 
particle packing which leads to disorder in the bond network. The macroscopic 
response of the model is mainly determined by the constitutive laws and 
breaking 
criteria of beams which have to be chosen to account for the observed 
materials' 
behavior. 

The primary fracture mechanism of cohesive frictional materials is the tensile 
and shear failure of bonds along the grain boundaries. This is captured by DEM 
approaches such that failure criteria of beams are formulated in terms of axial 
stresses and bending and torsion moments (strains) 
\cite{dem_basic_1,dem_basic_2,dem_basic_3,dem_basic_4,dem_basic_5}. Cracks form 
due to the gradual removal of cohesive elements as they fulfill the failure 
condition during the time evolution of the system. The structural disorder of 
the particle packing and bond network can be complemented by strength disorder 
treating the parameters of the failure criterion as stochastic variables 
\cite{humb1}.  Contact forces between particles are set on when cohesion is 
lost 
to prevent the penetration of crack faces into each other. The time evolution 
of 
the particle system is followed by molecular dynamics simulations, i.e.\ the 
equation of motion of all particles is solved numerically for the translational 
and rotational degrees of freedom with properly set initial, boundary, and 
loading conditions \cite{allen_til,grandynbook}. The model construction is 
illustrated in Fig.\ \ref{fig:dem} where a sedimentation algorithm was used to 
generate the initial particle packing (Fig.\ \ref{fig:dem}$(a)$)  
\cite{kun_rup_1,kun_rup_2}. Delaunay partitioning was carried out with the 
particle centers and beams were introduced between particles along the edges of 
tetrahedra (Fig.\ \ref{fig:dem}$(b)$). Finally, the model was applied to 
investigate the impact induced breakup of a rectangular specimen (Fig.\ 
\ref{fig:dem}$(c)$). 

\subsection{ Concurrent Discrete/Finite Element coupling}
Fracture in heterogeneous materials can also be understood by the flow of 
elastic energy from a volume into the formation of new internal surfaces 
\cite{F1}. Crack growth is thus a localization phenomenon with a spatially 
limited process zone. Physical access to the dynamic processes inside this zone 
can be obtained on a mesoscopic level by DEM simulation with a sufficient 
number 
of elements. However the process zone is embedded in an elastic foundation and 
usually the majority of particles is needed for representing the elastic 
domain, 
a job that can much more efficiently be dealt with by continuum methods. The 
last decade has seen an avalanche of works on different multi-scale methods for 
all kinds of applications and methods that can be classified to be either 
hierarchical or concurrent. 
\begin{figure}
\centering{\includegraphics[scale=0.7]{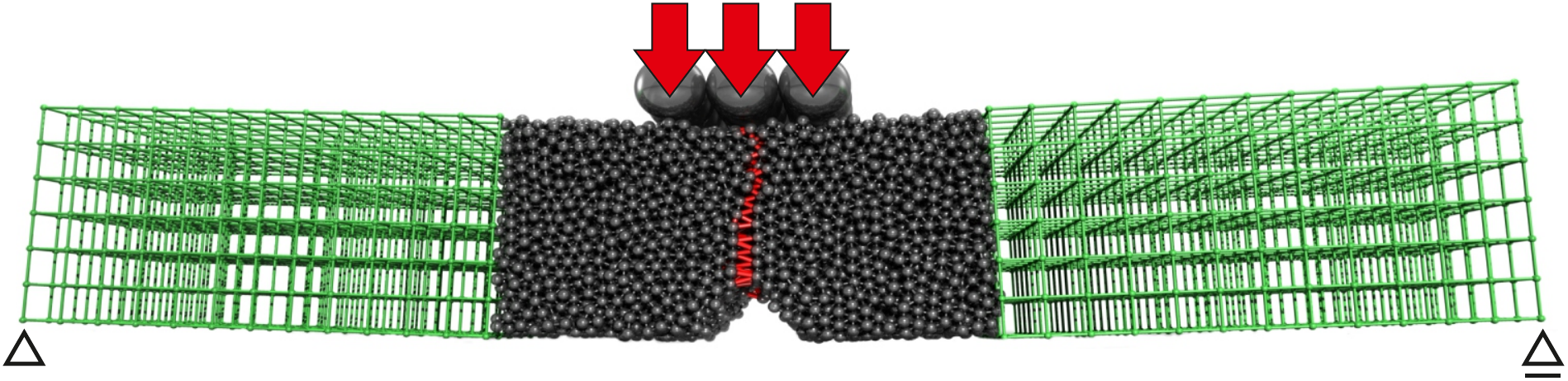}}
 \caption{Snapshot of a Charpy test with master-slave coupling of 
non-coincident 
nodes. Red elements resemble broken beams, green lines outline the edges of the 
20 node quadratic brick elements. \label{fig:fig1}}
\end{figure} 
The latter ones embrace all approaches where a fine-scale model is embedded and 
intimately coupled to a coarse-scale model like the example shown in Fig.\ 
\ref{fig:fig1}. For a comprehensive review on the methods, we refer to 
\cite{F2}. As we calculate dynamic interactions in the DEM, the challenge is to 
couple the DEM domain to the continuum domain in a way that the interface is 
without spurious reflections, or it other words "mechanically transparent". 
Since both methods discretize time and space, they can only resolve oscillations 
up to a cutoff frequency $w_{max}$ defined by the ratio of the wave speed with 
respect to the minimum node distance or characteristic particle size, 
respectively \cite{F2}. In general the cutoff 
frequency of the continuum domain $w_{max}^c >> w_{max}^{DEM}$ in order to 
benefit from a continuum approach. Unfortunately, this results in phonon 
reflections at the model interface for frequencies below $w_{max}^c$ that need 
to be mitigated in one way or the other.

\begin{figure}
\centering{\includegraphics[scale=0.7]{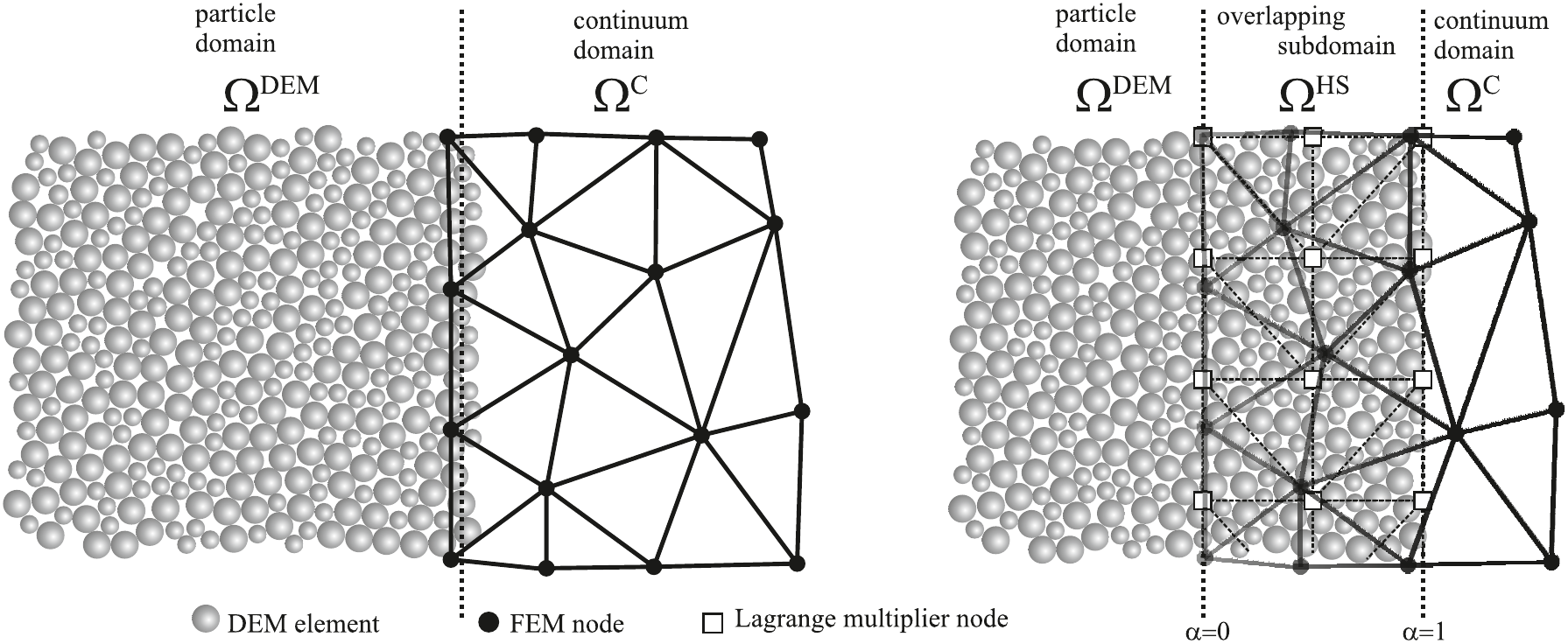}}
\caption{Edge-to-edge coupling of particle and continuum domain (left) and 
overlapping domain method with Lagrange multiplier mesh (right). 
\label{fig:fig2}}
\end{figure}
In principle one can impose a direct edge-to-edge or master-slave coupling and 
damp the reflected phonons close to the model interface in the DEM domain to 
obtain "silent boundaries" (see Fig.\ \ref{fig:fig1}). This master-slave 
coupling is a standard technique in FEM \cite{F3} and compatibility is enforced 
for all coupled degrees of freedom by constraining and mapping the slave DEM 
nodes onto the respective FEM master surface positions by the shape functions 
of 
the used elements. The forces and moments from the DEM nodes are in return 
added 
to the continuum model by standard contact procedures. Alternatively one can 
impose a smooth transition between models with an overlapping or bridging 
domain. The bridging domain method, proposed by Belytschko and Xiao \cite{F4} 
avoids sharp interphases by enforcing compatibility inside an overlapping 
domain 
by Lagrange multiplies.  Both methods are schematized in Fig.\ \ref{fig:fig2}.  
The linear scaling of relative importance of energy contributions of the 
different domains in the overlapping one by the blending function $\alpha$ 
assures a smooth transition between the domains.

\begin{figure}
\centering{\includegraphics[scale=0.7]{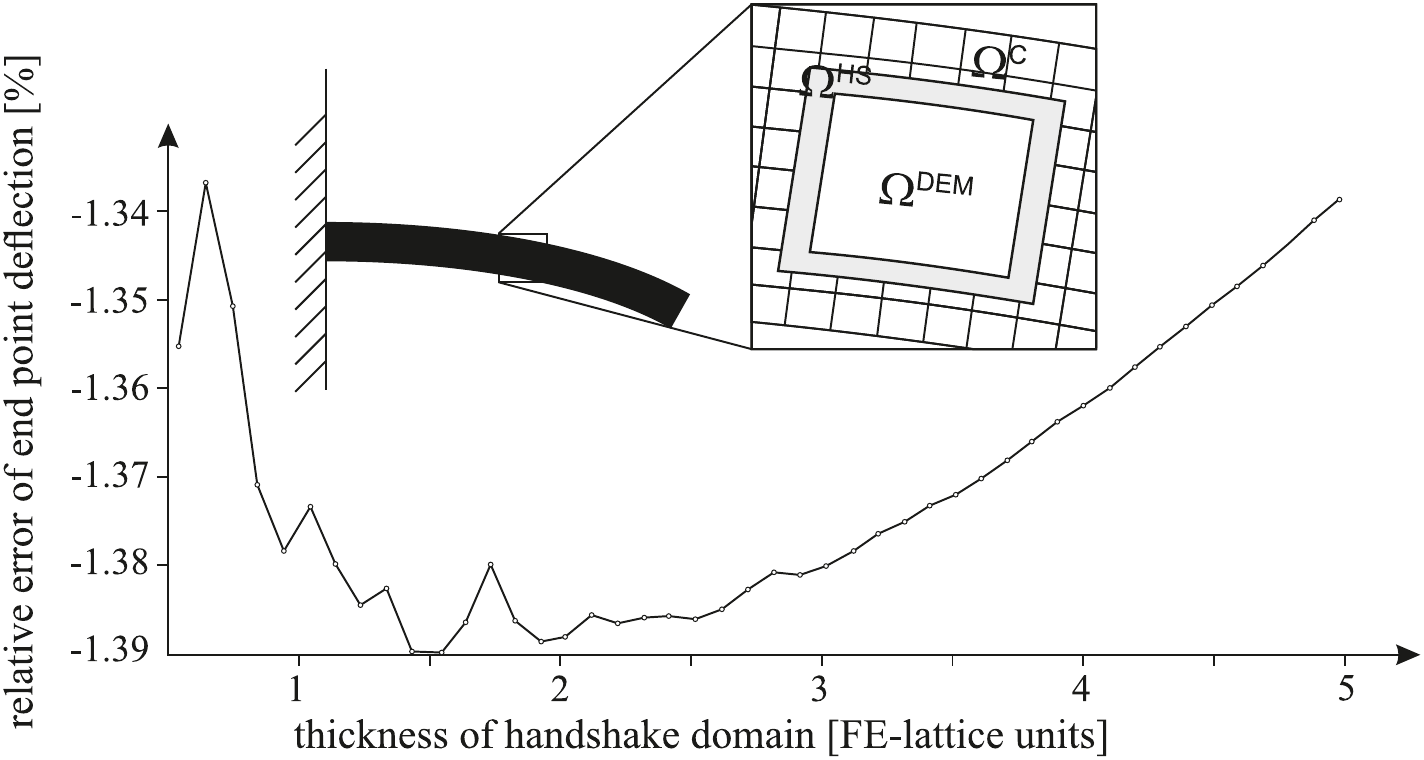}}
 \caption{Effect of relative size of handshake domain on accuracy of a  
cantilever beam problem with embedded particle mesh.\label{fig:fig3}}
\end{figure}

The shape functions of the Lagrange multiplier mesh, as well as element types 
and geometries can differ from those of the continuum mesh. In the simplest 
case 
linear functions on triangular elements that mesh the overlapping domain are 
chosen with linear blending functions, but Dirac delta functions and higher 
order blending functions are reported to work best \cite{F2}. The reason is 
that 
these strict Lagrange multipliers enforce exact compatibility with the finite 
element approximation and therefore fulfill the patch test, while other types 
of 
interpolations via shape functions donâ€™t and ghost forces exist. To assure a 
smooth transition, also the extension of the Lagrange multiplier field should 
be 
chosen such, that several FE nodes are captured (see Fig.\ \ref{fig:fig3}).

\section{Rupture cascades in the discrete element model}
Since the end of the '70s DEM gained widespread applications and had a 
substantial impact on our understanding of fracture processes of heterogeneous 
materials. In the following we highlight two aspects of breakdown phenomena 
where the heterogeneity of materials plays a crucial role and recently DEM 
simulations combined with the approach of statistical physics led to new 
understanding. First, we present how the statistics and dynamics of crackling 
noise emerging in slowly compressed brittle materials can be investigated in 
the 
framework of DEM, then we focus on the DEM modelling of fragmentation processes 
induced by energetic loading.

\subsection{Crackling noise during compressive failure}
Macroscopic failure of heterogeneous materials under slow external driving, 
i.e.\ under slowly increasing deformation or force, occurs as the culmination 
of 
damage accumulation \cite{herrmann}: at the beginning of the loading process 
micro-cracks nucleate at the weakest points in an uncorrelated way. Later on as 
the local stress fields of such defects interact, spatial correlation develops, 
which leads to growth of existing cracks and to an enhanced nucleation in their 
vicinity. The final stage of the process is dominated by the merging of cracks 
leading to the emergence of a macroscopic crack which spans the entire sample. 
However, this damage accumulation is not a "smooth" process, it proceeds in 
bursts of cracking events on the micro and meso scales. Such intermittent 
breaking avalanches generate elastic waves which can be recorded in the form of 
acoustic noise \cite{herrmann,eliz,mikko}. The acoustic emission technique is 
one of the most important diagnostic tools providing very valuable information 
about the microscopic dynamics of fracture \cite{osvany}. The recent progress 
achieved in experimental techniques addressed the question whether crackling 
noise measurements could be used to forecast the imminent catastrophic failure 
event. The problem has a high importance for the safety assessment of 
engineering constructions and for the forecasting of natural catastrophes such 
as landslides and
earthquakes \cite{bvalue1,bvalue2,bvalue3}.

Statistics of breaking bursts is usually investigated in the framework of 
stochastic lattice models, which are based on regular lattices of springs, 
beams, fibers, or fuses \cite{herrmann,mikko}. They have the advantage of 
allowing for a straightforward identification of breaking avalanches, however, 
they impose simplifications on the micro-structure of materials and on the 
dynamics of local breakings. Stochastic lattice models have qualitatively 
reproduced the integrated power law statistics of crackling noise and revealed 
interesting effects of the amount of disorder, friction, and range of load 
redistribution on the value of the exponent \cite{mikko,hidalgo,zapperi_size}. 
Both, under field conditions and in engineering applications, materials are 
often subject to compressive loading. Hence, the computational modelling of 
crackling noise under compression have a high practical importance, however, 
in this case lattice models of fracture face difficulties to fully capture 
the relevant microscopic mechanisms. To overcome this problem, recently, 
a DEM approach has been proposed 
to investigate the dynamics and statistics of rupture cascades 
\cite{kun_rup_1,kun_rup_2}. 

\subsubsection{Cascades of beam breaking}
In DEM crackling bursts are identified as cascades of micro-fractures, i.e.\  
correlated trails of breaking particle contacts which makes it possible to 
decompose the process of damage accumulation into a time series of elementary 
events of fracturing \cite{kun_rup_1,kun_rup_2}. 
\begin{figure}
\centering{\includegraphics[scale=0.5]{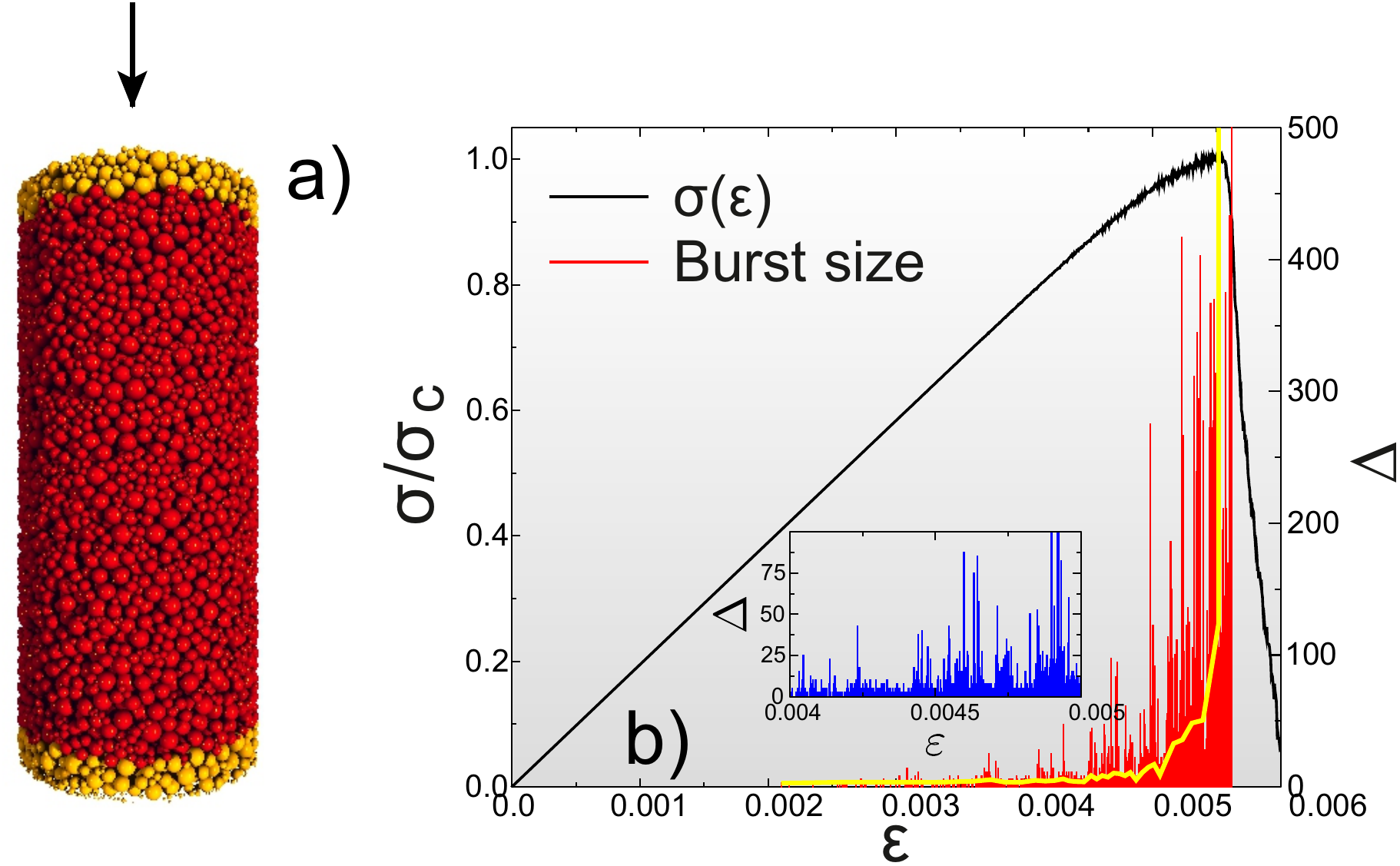}}
 \caption{$(a)$ Uniaxial loading of a cylindrical specimen of $20000$ particles 
was carried out in such a way that a few particle layers on the top and bottom 
were clamped (gold color) and the top layers were moving downward at a constant 
speed while the bottom was fixed. $(b)$ Constitutive curve 
$\sigma(\varepsilon)$ 
together with the sequence of breaking bursts in a single simulation. The burst 
size $\Delta$ is plotted at the strain $\varepsilon$  where the burst occurred. 
The yellow line indicates the moving average of burst sizes calculated over 50 
consecutive events.  The inset presents a magnified view on a smaller segment 
of 
the time series. \label{fig:constit}}
\end{figure} 
In these studies strain controlled uniaxial compression of cylindrical samples 
was simulated (see Fig.\ \ref{fig:constit}) measuring the macroscopic response 
of the system and the microscopic evolution of damage. A representative example 
of the constitutive curve $\sigma(\varepsilon)$ of the system is shown in Fig.\ 
\ref{fig:constit}$(b)$ where a quasi-brittle behavior is evidenced. Simulations 
revealed that in spite of the smooth macroscopic response, on the micro-scale 
the accumulation of damage, i.e.\ the breaking of beam elements proceeds in a 
jerky way. The reason is that after a beam breaks, the released stress must get 
redistributed in the surrounding volume. It enhances the load on neighboring 
beams which may induce further breakings and in turn it can even trigger an 
entire avalanche of breakings. In a DEM framework the breaking sequence of 
beams 
can be traced by recording the time $t^b_i$ and position $\vec{r}^b_i$ of 
single 
breakings. In order to quantify the temporal clustering of breaking events it 
is 
assumed that consecutive breaking events are correlated if they follow each 
other within a correlation time $t_c$, i.e.\ if the condition 
$t^b_{i+1}-t^b_i<t_c$ is fulfilled. The value of the correlation time $t_c$ can 
be physically motivated, namely, it is the time needed for the stress release 
wave to cross the specimen.

Based on the concept of correlated breakings, bursts of local failure events 
can 
be identified. The burst size $\Delta$ is the number of beams breaking in the 
avalanche which is related to the new crack surface created by the burst. It 
can 
be observed in Fig.\ \ref{fig:constit}$(b)$ that during the loading process the 
size of bursts $\Delta$ has strong fluctuations due to the quenched structural 
disorder of the material but its average has an increasing tendency towards 
failure. This generic behavior is in a nice qualitative agreement with the 
outcomes of acoustic emission measurement on heterogeneous materials 
\cite{mikko,osvany,bvalue1,bvalue2,bvalue3}. In the framework of DEM, further 
useful quantities can be defined to characterize single crackling avalanches 
and 
the evolution of their time series: Besides the burst size $\Delta$, the time 
of 
occurrence $t$ and duration $T$ are of particular interest together with the 
amount of energy $E$ released by bursts. The temporal sequence of avalanches 
can 
be characterized by the waiting time $t_w$ between consecutive events. 
\begin{figure}
\centering{\includegraphics[scale=0.7]{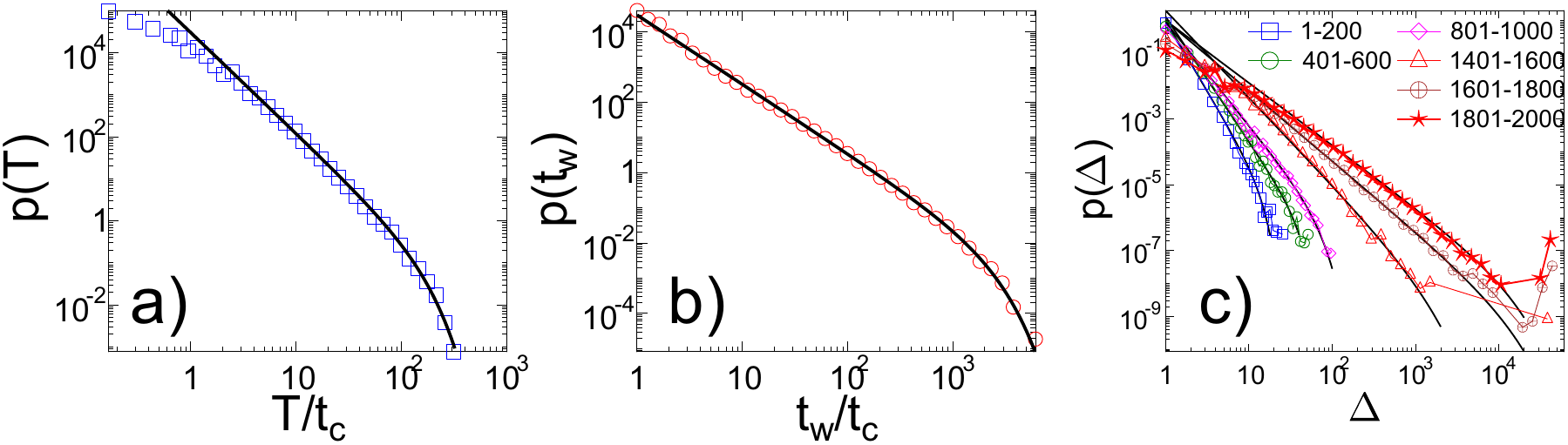}}
 \caption{Probability distribution of the duration $p(T)$ of bursts $(a)$ and 
of 
waiting times $p(t_w)$ $(b)$ between consecutive events averaged over 800 
samples.  $(c)$ The size distribution of bursts $p(\Delta)$ calculated in 
windows of 200 events. The legend indicates which events are included in the 
analysis. The continuous black lines represent fits with Eq.\ 
(\ref{eq:power_dist}).\label{fig:crackling}}
\end{figure} 
\subsubsection{Statistics of crackling events}
The integrated statistics of the characteristic quantities, i.e.\ the 
probability distributions of the burst size $\Delta$, energy $E$, and duration 
$T$, furthermore, of the waiting time $t_w$ -- considering all events up to 
failure -- proved to have a power law functional form with a stretched 
exponential cutoff
\begin{equation}\label{eq:power_dist}
p(x) \sim x^{-\alpha} \exp{\left[-(x/x^*)^c\right]}.
\end{equation}
Here $x$ is a generic notation for $\Delta$, $E$, $T$, and $t_w$, while $x^*$ stands
for the characteristic scale of the corresponding quantity. Representative 
examples are shown in Figs.\ \ref{fig:crackling}$(a)$ and $(b)$ for the 
distributions of burst durations $p(T)$ and waiting times $p(t_w)$, where the 
continuous lines represent high quality fits with Eq.\ (\ref{eq:power_dist}). 
The results of DEM simulations \cite{kun_rup_1,kun_rup_2} have an excellent 
agreement with the experimental findings on the statistics of acoustic bursts 
accompanying the compressive failure of sedimentary rocks such as sand stones 
\cite{bvalue1,bvalue2,bvalue3}.

Based on the detailed information DEM provides about the evolution of the 
crackling time series, it is also possible to investigate how the statistics of 
crackling events changes as the system approaches macroscopic failure. Figure 
\ref{fig:crackling}$(c)$ demonstrates size distributions $p(\Delta)$ 
considering 
bursts in windows of 200 consecutive events instead of the integrated 
statistics. For all curves the functional form of Eq.\ (\ref{eq:power_dist}) is 
evidenced, however, the value of the exponent of the power law regime decreases 
from 4.25 to 1.5 when approaching macroscopic failure 
\cite{kun_rup_1,kun_rup_2}. This behavior is in an excellent qualitative 
agreement with the so-called "b-value" anomaly observed for earthquakes and in 
laboratory experiments on compressive fracture of rocks, i.e.\ the exponent $b$ 
of the magnitude distribution of crackling events decreases when approaching 
the 
critical point of global failure \cite{bvalue1,bvalue2,bvalue3}. 

Recent simulations also demonstrated the potential of DEM to investigate the 
spatial structure of damage, the gradual emergence of spatial correlation of 
consecutive events, the formation of the damage band due to the dominance of 
shear in the failure process, and even the gradual fragmentation of pieces in 
the damage band \cite{kun_rup_1,kun_rup_2}.

\subsection{Fragmentation phenomena}
Energetic loading leads to fragmentation with a multitude of cracks forming 
simultaneously. This leads to a rapid disintegration of solids into a large 
number of pieces. On a longer time scale repeated loading or shearing under a 
high pressure give rise to a similar outcome with fragment sizes spanning a 
broad range with a scale free probability distribution 
\cite{frag_rev1,frag_rev2,turcotte}. In Nature fragmentation of solid bodies 
occur on a broad range of length and time scales from the collision induced 
breakup of asteroids down to the degradation processes in a fault gauge. 
Detailed knowledge on fragmentation is required in the industry where it is 
exploited by technologies of mining and ore processing.  In particular such 
applied but also fundamental questions on fragmentation processes are most 
suitably answered by DEM simulations.

\subsubsection{Universality in fragmentation}
\begin{figure}
\centering{\includegraphics[scale=0.65]{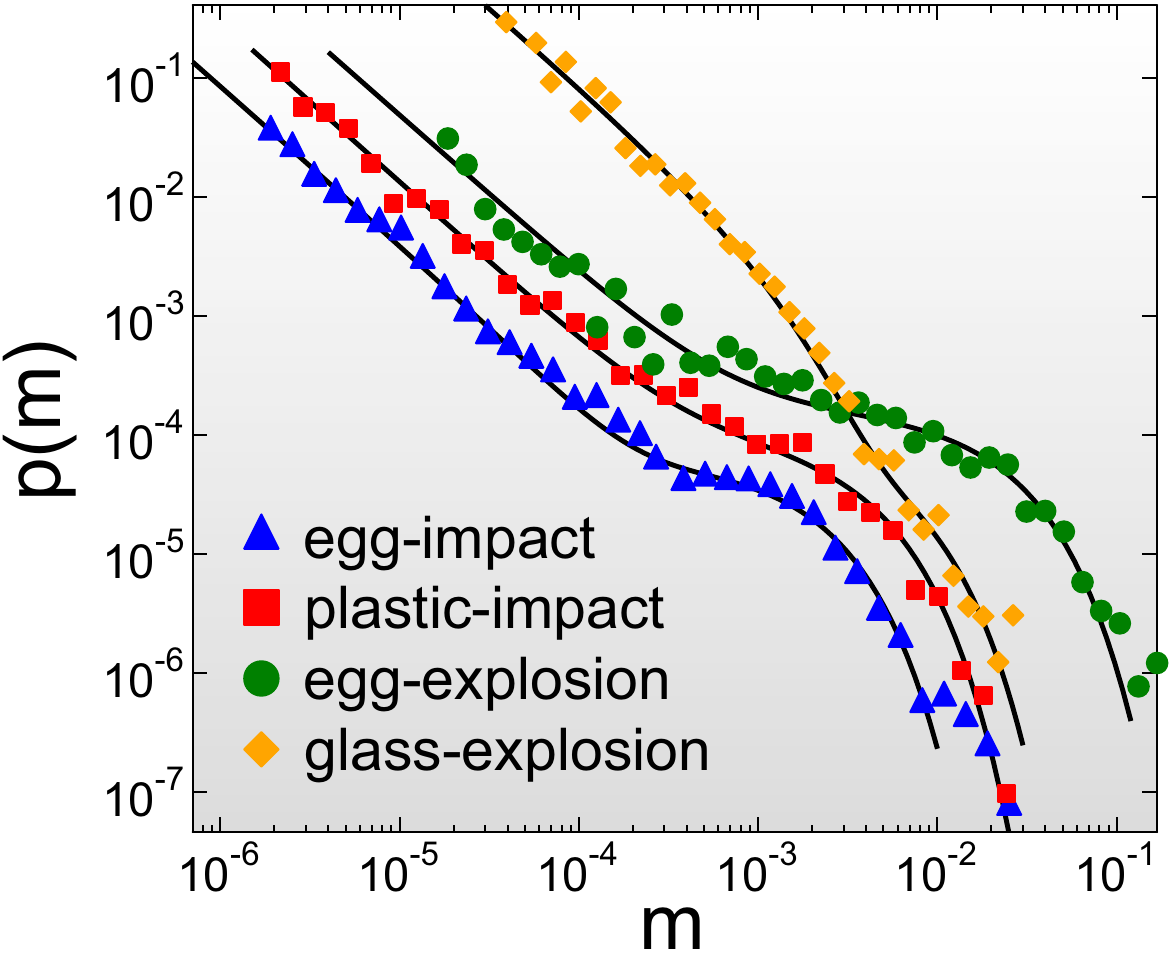}}
 \caption{Universality of fragment mass distributions of shell systems for 
different materials and types of energetic loading: brittle eggshells were 
fragmented both by explosion and impact against a hard wall. Glass balls were 
exploded, while plastic shells were impacted to a hard wall after making them 
brittle at the temperature of liquid nitrogen \cite{wittel_frag1}.  The 
continuous line represent fits with Eq.\ (\ref{eq:fragmass}) such that the 
value 
of $\tau$ is $1.35$ in all cases showing the universality. 
\label{fig:fragment_univ}}
\end{figure}

The most remarkable feature of fragmentation phenomena is that the value of the 
power law exponent of the size distribution of pieces shows an astonishing 
robustness being independent of the way of loading, of material properties, and 
relevant length scales. During the past decades the understanding of the 
observed universality has been the main driving force of fragmentation 
research. 
Experimental and numerical investigations have revealed that the universality 
classes of fragmentation phenomena are mainly determined by the dimensionality 
of the system \cite{frag_rev1,frag_rev2,turcotte,astrom2} and by the 
brittle/ductile character of the mechanical response of the material 
\cite{timar1}. For brittle materials the underlying breakup mechanisms 
originate 
from crack tip instabilities that lead to repeated crack branching-merging 
\cite{astrom2}. Combining the branching-merging scenario with the Poissonian 
nature of the initial nucleation of major cracks a complex functional form was 
proposed which describes the complete mass/size distribution of fragments 
$p(m)$ 
including the cutoff regime, as well \cite{astrom2}
\begin{equation}
p(m) ~ \sim (1-\beta)m^{-\tau}\exp{(-m/\overline{m}_0)} +
\beta\exp{(-m/\overline{m}_1)}. 
\label{eq:fragmass}
\end{equation}
Here, $\tau$ denotes the exponent of the power law regime, $\overline{m}_0$ and 
$\overline{m}_1$ are characteristic fragment masses, and $\beta$ controlls the  
contribution of the two terms of the right hand side \cite{astrom2}. The 
universality of fragment mass/size distributions is demonstrated in Fig.\ 
\ref{fig:fragment_univ} for closed shells in 3D where shells made of three 
different materials were fragmented by explosion and impact against a hard 
wall. 
In the regime of small fragment masses best fit was obtained with Eq.\ 
(\ref{eq:fragmass}) using a unique exponent $\tau=1.35\pm 0.02$ which defines 
the universality class of brittle shells.
High speed imaging of shell fragmentation provided direct proof of the 
predicted 
breaking scenario \cite{wittel_frag1}, and additionally it revealed that not 
only fragment sizes but even the shape of fragments obeys scaling laws 
\cite{kun_frag3}.

\subsubsection{DEM simulations of fragmentation processes}
Due to the difficulties of the experimental investigation of fragmentation 
phenomena, DEM simulations had a major contribution to the development of 
the field. Although, recent applications of high-speed cameras and 3D imaging 
have allowed for a deeper insight into the process of the rapid breakup of solids 
\cite{khanal,chau,kadono,kun_frag3}, 
DEM simulations still complement the experiments with very valuable information.
Energetic loading like explosion or impact generate a large number of 
simultaneously growing cracks which interact with each other in a complicated 
way. DEM has the capabilities to handle this high degree of complexity making 
possible
a realistic treatment of fragmentation processes.

Detailed studies with DEM in various embedding dimensions revealed a transition 
from damage to fragmentation \cite{kun_frag1} at a critical imparted energy 
already for two-dimensional systems. The existence of the damage-fragmentation 
critical point has been confirmed by further DEM simulations 
\cite{timar2,myagkov} of various types of fragmentation processes, and it was 
also reproduced by experiments \cite{katsuragi,katsuragi1}. The result implies 
that 
universality of fragment size distributions is due to the underlying continuous 
phase transition. The entire richness of fragmentation mechanisms however could 
only be resolved by full 3D systems once a significant particle number could be 
considered \cite{humb1}. Contrary to the simple branching-merging scenario it 
became clear that there exist different mechanisms which get activated as the 
imparted energy increases and their interaction determines the final breaking 
scenario \cite{humb1}. 
\begin{figure}
\centering{\includegraphics[scale=0.7]{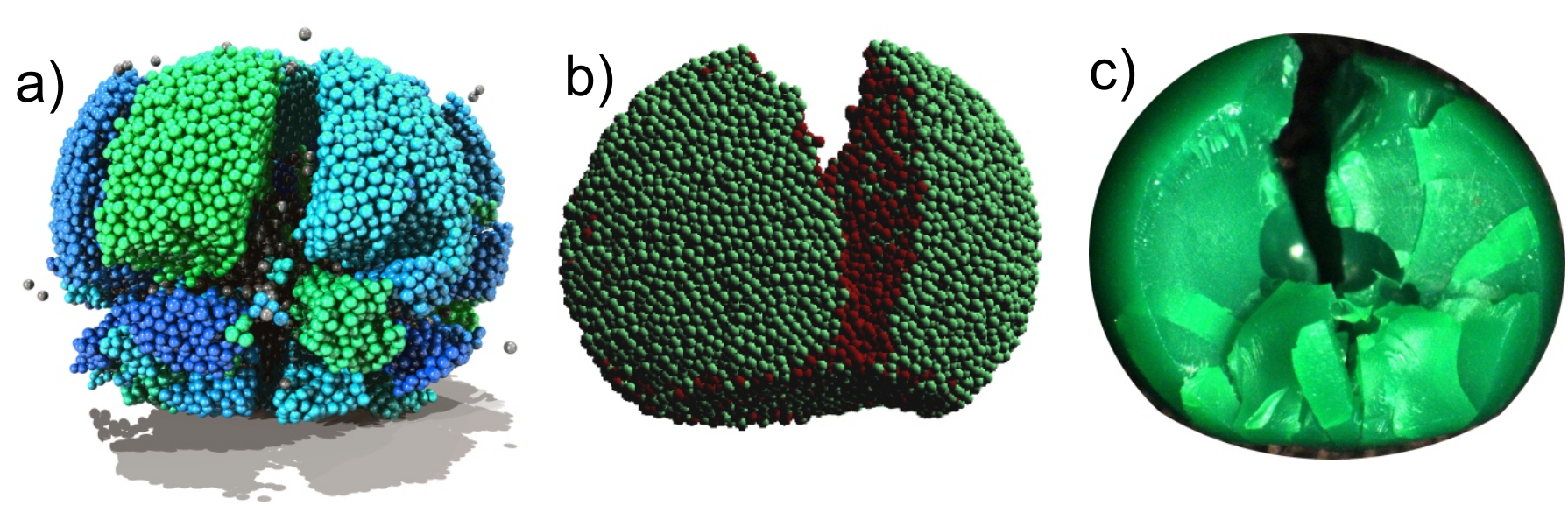}}
 \caption{$(a)$ Discrete element simulation of the fragmentation of a brittle 
sphere induced by impact against a hard wall \cite{humb1}. The impact velocity 
falls slightly above the critical point of the damage-fragmentation transition. 
$(b)$ In a similar impact simulation of a plastically deforming sphere at low 
impact velocities, a single crack occurs in the middle and a large permanent 
deformation remains around the impact site \cite{timar1}. $(c)$ Impact 
experiments of plastic balls revealed a similar breakup mechanism in a good 
quantitative agreement with DEM simulations \cite{timar1}.  
\label{fig:fragment_dem}}
\end{figure} 
Fragmentation processes of plastically deforming materials show an even higher 
complexity: power law size distribution of fragments has been confirmed by 
experiments, however, with an exponent significantly lower than for brittle 
materials. DEM simulations clarified that shear induced breaking is responsible 
for the emergence of the novel universality class which makes plastic 
fragmentation similar to the one of liquid droplets \cite{timar1}. The effect 
of 
material microstructures on the outcome of the fragmentation was studied by 
mapping material micro-structures of multi-phase materials and composites onto 
the DEM systems \cite{humb2}. Surprisingly the size distribution exponent is 
rather robust with respect to such issues, strengthening the universal 
character 
of fragmentation. Representative examples of DEM simulations of fragmentation 
processes are presented in Fig.\ \ref{fig:fragment_dem} both for brittle and 
ductile materials. The figure also demonstrates the agreement of simulations 
with experiments. 

\section{Conclusions and future challenges}
In the present paper we briefly reviewed the basic ideas behind DEM for 
heterogeneous materials and highlighted two fields where this modelling 
approach 
played a decisive role to reach recent success. Profiting from the  increase of 
computer power and the success of hybridization of modelling approaches 
simulation studies of fracture and fragmentation phenomena can help to resolve 
current debates of the field and to reach new challenges. When studying 
statistical features of the fracture of heterogeneous materials such as the 
size 
effect of macroscopic strength and crackling noise generated by avalanches of 
micro-fractures, stochastic lattice models have been successfully applied under 
tensile loading conditions \cite{zapperi_size}. 
However, under compressive loading they usually have 
difficulties to account for all relevant mechanisms. We have demonstrated that 
DEM offers an adequate modelling framework for crackling phenomena under 
compression reproducing all observed scaling laws of rupture cascades obtained 
by field measurements, as well as, in laboratory experiments 
\cite{kun_rup_1,kun_rup_2}. The results imply that DEM has a high potential to 
understand the emergence of catastrophic failure in porous granular media 
challenging Earth sciences and engineering. In natural catastrophes such as 
landslides and earthquakes, the available data are often incomplete and provide 
only a limited insight into the complexity of processes that lead to failure. 
The main contribution of DEM is that it can capture all relevant processes down 
to the length scale of single grains, and hence, it can reveal mechanisms 
hidden 
for experimental approaches. With these capabilities for the investigation of 
the statistics and dynamics of rupture cascades DEM may give rise to a 
breakthrough in developing predictive models of catastrophic failures and 
even earthquakes in the near future.

Research on fragmentation faces similar challenges. Recent experiments on 
impact 
induced fragmentation of one- and two-dimensional objects revealed that the 
power law exponent $\tau$ of the fragment size/mass distribution increases with 
the imparted energy \cite{ching}. DEM simulations performed in two dimensions 
confirmed this finding and yielded a logarithmic dependence of $\tau$ on the 
energy \cite{sator1,sator2}. The results are remarkable because on the one hand 
they question universality and the underlying phase transition picture of the 
damage-fragmentation transition, and on the other hand, they have relevance for 
industrial applications in mining and ore processing, as well. 
However, recent DEM studies on the breakup of 
spherical bodies due to impact against a hard wall demonstrated that the 
apparent increase of the exponent can be removed by rescaling the mass/size 
distributions with the average fragment mass \cite{timar2}. Finite size scaling 
proved to be indispensable to correctly determine characteristic exponents of 
fragmentation phenomena which again shows the importance of large system sizes 
and calls for further investigations to settle the problem. Both experimental 
and theoretical investigations have shown that the dimensionality of the 
breakup process, especially the interplay of the dimensionality of the object 
and of the 
embedding space, plays a crucial role in the selection of the dominating 
mechanism of dynamic cracking and fragment formation 
\cite{wittel_frag1,astrom2,astrom3}. This addresses the opportunity that in 
certain cases universality can be violated and the energy dependence can be 
understood through the gradual activation of different breakup mechanism and 
the mixing of them as the imparted energy is varied. 

Comminution of solid particles in ore processing is one of the most important 
applications of fragmentation inducing a huge cost for the industry. Improving
the efficiency of size reduction in comminution machinery and the separation 
of fragments according to their size are still important driving forces of 
fragmentation research greatly profiting from DEM.

Advancement of measuring technologies has made it possibly to go beyond the 
analysis of the mass/size distribution of fragments in the final state of the 
breakup process. There is an increasing amount of information available on the 
velocity of pieces, as well. It is a great challenge for theoretical 
investigations to understand what determines the functional form of the 
velocity 
distribution of fragments, and whether the mass and velocity of fragments are 
correlated. Beyond their scientific importance both problems have also 
practical 
relevance: on orbit fragmentation events are the main source of space debris 
where the velocity of debris pieces and the presence of mass-velocity 
correlation are crucial for estimating the risk of damaging collisions with 
satellites. 

Today particle models for fracture and fragmentation are at the verge of 
becoming significant tools for simulating industrial processes. The dilemma of 
either using a large number of spherical particles or a significantly smaller 
number of aggregated or polygonal particles for discretization is slowly 
diluted 
by the development of computer hardware and should vanish within the next 
decade. Additionally, the incorporation of DEM into FEM workbenches will bring 
these methods to a wider community of applied users. With every new release of 
FEM simulation suites, software companies extend functionality, recently to 
incorporate particle methods. Even though still the simplest methods are 
implemented, soon we might see advanced DEM embedded in FEM code with robust 
concurrent coupling. As the continuum and discrete worlds continuously merge 
inside commercial software packages users are increasingly liberated from 
technicalities of discretization and implementation issues. 
Till today a major problem of engineering design is the fracture size effect, i.e.\ 
the fracture strength of heterogeneous materials decreases with increasing sample size 
\cite{zapperi_size}.
The size effect is the consequence of the complex interplay of the long range elastic 
interaction of material elements and of the inherent disorder. Since DEM naturally
accounts for both, with the increase of computer power and algorithmic developments 
DEM is expected to have important contribution to tackle fracture size effect.

\section*{Acknowledgements}
{\it This work is devoted to our mentor and friend Hans J. Herrmann for his 
60th birthday - a researcher that stimulated a multitude of innovations in this 
field.}


\begin{thebibliography}{99}

\bibitem{herrmann} H.J.\ Herrmann and S.\ Roux (Edts.), \textit{Statistical 
Models for the Fracture of Disordered media} (Elsevier Science Publishers, 
Amsterdam, 1990).
\bibitem{eliz} E.\ Bouchaud, D.\ Jeulin, C.\ Prioul, and S.\ Roux (Edts), 
\textit{Physical Aspects of Fracture} (Kluwer Academic Publishers, New York, 
2001).
\bibitem{mikko} M.J.\ Alava, P.\ K.\ V.\ V.\ Nukala,  and S.\ Zapperi, Adv.\ 
Phys.\ {\bf 55}, (2006) 349.
\bibitem{dem_basic_1} P.\ A.\ Cundall, O.\ D.\ Strack, Geotechnique {\bf 29}, 
(1979) 47.
\bibitem{allen_til} M.\ P.\ Allen and D.\ J.\ Tildesley, \textit{Comuter 
simulation of liquids} (Oxford University Press, Oxford, 1984).
\bibitem{sph}  L.\ D.\ Libersky and A.\ G.\ Petschek, Lecture Notes in Physics 
{\bf 395}, (1990) 248.
\bibitem{F1} A.\ T.\ Zehnder, \textit{Fracture Mechanics, Lecture notes in 
applied and computational mechanics} {\bf 62} (Springer, Berlin, 2012).
\bibitem{F2} T.\ Belytschko, R.\ Gracie, M.\ Xu, \textit{Concurrent Coupling of 
Atomistic and Continuum Models}, in J.\ Fisch (Edt.), \textit{Multiscale 
Methods: Bridging the Scales in Science and Engineering} (Oxford University 
Press, Oxford, 2010), 93.
\bibitem{F3} T.\ Belytschko, W.\ K.\ Liu, and B.\ Moran, \textit{Non-linear 
Finite Elements for Continua and Structures} (John Wiley \& Sons Inc., New 
York, 
2001).
\bibitem{F4} T.\ Belytschko and S.\ P.\ Xiao, Int.\ J.\ Mult.\ Comp.\ Engin.\ 
{\bf 1}, (2003) 115.
\bibitem{F5} A.\ Leonardi, F.\ K.\ Wittel, M.\ Mendoza, H.\ J.\ Herrmann, 
Comp.\ 
Part.\ Mech.\ 1 (1/2014).
\bibitem{dem_basic_2} E.\ Schlangen, E.\ J.\ Garboczi, Engrg.\ Fract.\ Mech.\ 
{\bf 57} (1997) 319.
\bibitem{dem_basic_3} D.\ Potyondy and P.\ Cundall, Int.\ J.\ Rock Mech.\ Min.\ 
Sci.\ {\bf 41}, (2004) 1329.
\bibitem{dem_basic_4} G.A.\ D'Addetta, F.\ Kun, E.\ Ramm, H.J.\ Herrmann, 
Lecture Notes in Physics {\bf 568}, (2001) 231.
\bibitem{dem_basic_5} G.\ A.\ D'Addetta, F.\ Kun, E.\ Ramm, Gran.\ Mat.\ {\bf 
4}, (2002) 77.
\bibitem{kun_basquin} F.\ Kun, H.\ A.\ Carmona, J.\ S.\ Andrade, and H.\ J.\ 
Herrmann, Phys.\ Rev.\ Lett.\ {\bf 100}, (2008) 094301.
\bibitem{humb1} H.\ A.\ Carmona, F.\ K.\ Wittel, F.\ Kun, and H.\ J.\ Herrmann, 
Phys.\ Rev.\ E {\bf 77}, (2008) 051302.
\bibitem{grandynbook} T.\ P\"oschel and T.\ Schwager, \textit{Computational 
Granular Dynamics} (Springer, Berlin, 2005).
\bibitem{kun_rup_1} F. Kun, I. Varga, S. Lennartz-Sassinek, and I. G. Main, 
Phys.\ Rev.\ Lett.\ {\bf 112},
(2014) 065501.
\bibitem{kun_rup_2} F. Kun, I. Varga, S. Lennartz-Sassinek, and I. G. Main, 
Phys.\ Rev.\ E {\bf 88},
(2013) 062207.
\bibitem{geomalg} W.\ Salvat, N.\ Mariani, G.\ Barreto, O.\ Martinez, Catal.\ 
Today {\bf 107â€“108}, (2005) 513.
\bibitem{bagi1} K.\ Bagi, Gran.\ Mat.\ {\bf 7}, (2005) 31.
\bibitem{donze1} J.-F.\ Jerier, D.\ Imbault, F.-V.\ Donze, and P.\ Doremus, 
Gran.\ Mat.\ {\bf 11}, (2009) 43.
\bibitem{donze2} S.\ Hentz, F.\ V.\ Donze, and L.\ Daudeville, Computers and 
Structures {\bf 82} (2004) 2509.
\bibitem{feng1} Y.\ T.\ Feng, D.R.J.\ Owen, Int.\ J.\ Numer.\ Meth.\ Engng.\ 
{\bf 56}, (2003) 699.
\bibitem{cui1} L.\ Cui and C.\ Oâ€™Sullivan, Gran.\ Matt.\ {\bf 5}, (2003) 135.
\bibitem{luding1} S.\ Luding, Gran.\ Mat.\ {\bf 10}, (2008) 235.
\bibitem{thornton} B.\ K.\ Mishra and C.\ Thornton, Int.\ J.\ Min.\ Process.\ 
{\bf 61}, (2001) 225.
\bibitem{osvany} M.\ Stojanova, S.\ Santucci, L.\ Vanel, and O.\ Ramos, Phys.\ 
Rev.\ Lett.\  {\bf 112}, (2014) 115502.
\bibitem{bvalue1} P.\ R.\ Sammonds, P.\ G.\ Meredith, and I.\ G. Main, Nature 
{\bf 359}, (1992) 228.
\bibitem{bvalue2} C.\ G.\ Hatton, I.\ G.\ Main, and P.\ G.\ Meredith, J.\ 
Struct.\ Geol.\ {\bf 15}, (1993) 1485.
\bibitem{bvalue3} I.\ Ojala, B.\ T.\ Ngwenya, I.\ G.\ Main, and S.\ C.\ 
Elphick, 
J.\ Geophys.\ Res.\ {\bf 108}, (2003) 2268.
\bibitem{hidalgo} R.\ C.\ Hidalgo, F.\ Kun, K.\ Kov\'acs, and I.\ 
Pagonabarraga, 
Phys.\ Rev.\ E {\bf 80}, (2009) 051108.
\bibitem{frag_rev1} H.\ J.\ Herrmann, F.\ K.\ Wittel, and F.\ Kun, Physica A 
{\bf 371}, (2006) 59. 
\bibitem{frag_rev2} J.\ Astr\"om, Adv.\ in Phys.\ {\bf 55}, (2006) 247.
\bibitem{turcotte} D.L.\ Turcotte, J.\ Geophys.\ Res.\ {\bf 91}, (1986) 1921.
\bibitem{khanal} M.\ Khanal, W.\ Schubert, and J.\ Tomas, Int.\ J.\ Miner.\ 
Process.\ 
{\bf 86}, (2008) 104.
\bibitem{chau} K.T.\ Chau, X.X.\ Wei, R.H.C.\ Wong, and T.X.\ Yu,
Mechanics of Materials {\bf 32}, (2000) 543.
\bibitem{kadono} T.\ Kadono and M.\ Arakawa, Phys.\ Rev.\ E {\bf 65}, (2002) 
035107(R).
\bibitem{kun_frag3} F.\ Kun, F.\ K.\ Wittel, H.\ J.\ Herrmann, B.-H.\ 
Kr\"oplin, 
and K.-J.\ Maloy, Phys.\ Rev.\ Lett.\ {\bf 96},  (2006) 025504.
\bibitem{astrom2} J.\ A.\ Astr\"om, F.\ Ouchterlony, R.\ P.\ Linna, and J.\ 
Timonen, Phys.\ Rev.\ Lett.\ {\bf 92}, (2004) 245506.
\bibitem{katsuragi} H.\ Katsuragi, D.\ Sugino, and H.\ Honjo, Phys.\ Rev.\ E 
{\bf 68}, (2003) 046105.
\bibitem{katsuragi1} H.\ Katsuragi, S.\ Ihara, and H.\ Honjo, 
Phys.\ Rev.\ Lett.\ {\bf 95}, (2005) 095503.
\bibitem{wittel_frag1} F. K. Wittel, F. Kun, H. J. Herrmann, and B.-H. Kroplin, 
Phys.\ Rev.\ Lett.\ {\bf 93}, (2004) 035504.
\bibitem{kun_frag1} F.\ Kun and H.\ J.\ Herrmann, Phys.\ Rev.\ E {\bf 59}, 
(1999) 2623. 
\bibitem{timar2} G.\ Tim\'ar, F.\ Kun, H.\ A.\ Carmona, and H.\ J.\ Herrmann, 
Phys.\ Rev.\ E {\bf 86}, (2012) 016113.
\bibitem{myagkov} N.\ N.\ Myagkov and T.\ A.\ Shumikhin, Physica A {\bf 358}, 
(2005) 423. 
\bibitem{timar1} G.\ Tim\'ar, J.\ Bl\"omer, F.\ Kun, and H.\ J.\ Herrmann, 
Phys.\ Rev.\ Lett.\ {\bf 104}, (2010) 095502.
\bibitem{humb2} H.\ A.\ Carmona, A.\ V.\ Guimaraes, J.\ S.\ Andrade Jr., I.\ 
Nikolakopoulos, F.\ K.\ Wittel, and H.\ J.\ Herrmann, under preparation.
\bibitem{ching} E.\ S.\ C.\ Ching, Y.\ Y.\ Yiu, K.\ F.\ Lo, Physica A {\bf 
265}, 
(1999) 119. 
\bibitem{sator1} N.\ Sator, S.\ Mechkov, and F.\ Sausset,  Europhys.\ Lett.\ 
{\bf 81}, (2008) 44002.
\bibitem{sator2} N.\ Sator and H.\ Hietala, Int.\ J.\ Fract.\ {\bf 163}, (2010) 
101.
\bibitem{astrom3} J.\ A.\ Astr\"om, Phys.\ Rev.\ E {\bf 80}, (2009) 046113.
\bibitem{zapperi_size} M.\ J.\ Alava, P.\ K.\ V.\ V.\ Nukala and S.\ Zapperi, 
J.\ Phys.\ D {\bf 42}, (2009) 214012. 

\end{thebibliography}
\end{document}